\documentclass[prl,showpacs,twocolumn,floatfix,superscriptaddress,amsmath,amssymb,10pt,aps]{revtex4-1}
\usepackage[utf8]{inputenc}
\usepackage[T1]{fontenc}
\usepackage{dsfont}
\usepackage{amsmath}
\usepackage{graphicx}
\usepackage{nicefrac}
\usepackage{xcolor}
\usepackage{natbib}
\usepackage{url}
\usepackage{ulem}

\begin{document}
\title{Spin Structure of K Valleys in Single-Layer WS$_2$ on Au(111)}

\author{Philipp Eickholt}
\email{philippeickholt@uni-muenster.de}
\affiliation{Physikalisches Institut, Westf\"alische Wilhelms-Universit\"at M\"unster, Wilhelm-Klemm-Stra{\ss}e 10, 48149 M\"unster, Germany}

\author{Charlotte Sanders}
\affiliation{Department of Physics and Astronomy, Interdisciplinary Nanoscience Center (iNANO), Aarhus University, 8000 Aarhus C, Denmark}

\author{Maciej Dendzik}
\affiliation{Department of Physics and Astronomy, Interdisciplinary Nanoscience Center (iNANO), Aarhus University, 8000 Aarhus C, Denmark}

\author{Luca Bignardi}
\affiliation{Elettra-Sincrotrone Trieste S.C.p.A., 34149 Trieste, Italy}

\author{Daniel Lizzit}
\affiliation{Elettra-Sincrotrone Trieste S.C.p.A., 34149 Trieste, Italy}

\author{Silvano Lizzit}
\affiliation{Elettra-Sincrotrone Trieste S.C.p.A., 34149 Trieste, Italy}

\author{Albert Bruix}
\affiliation{Department of Physics and Astronomy, Interdisciplinary Nanoscience Center (iNANO), Aarhus University, 8000 Aarhus C, Denmark}

\author{Philip Hofmann}
\affiliation{Department of Physics and Astronomy, Interdisciplinary Nanoscience Center (iNANO), Aarhus University, 8000 Aarhus C, Denmark}

\author{Markus Donath}
\affiliation{Physikalisches Institut, Westf\"alische Wilhelms-Universit\"at M\"unster, Wilhelm-Klemm-Stra{\ss}e 10, 48149 M\"unster, Germany}

\pacs{}

\begin{abstract}
The spin structure of the valence and conduction bands at the $\overline{\text{K}}$ and $\overline{\text{K}}$' valleys of single-layer WS$_2$ on Au(111) is determined by spin- and angle-resolved photoemission and inverse photoemission. The bands confining the direct band gap of 1.98 eV are out-of-plane spin polarized with spin-dependent energy splittings of 417 meV in the valence band and 16 meV in the conduction band. The sequence of the spin-split bands is the same in the valence and in the conduction bands and opposite at the $\overline{\text{K}}$ and the $\overline{\text{K}}$' high-symmetry points. The first observation explains "dark" excitons discussed in optical experiments, the latter points to coupled spin and valley physics in electron transport. The experimentally observed band dispersions are discussed along with band structure calculations for a freestanding single layer and for a single layer on Au(111).
\end{abstract}

\maketitle
Since the discovery of graphene, two-dimensional materials have driven intense research effort due to their fascinating electronic and optical properties \cite{Wang.2012}. The option of stacking different two-dimensional materials on top of each other opens the way of tailoring specific material properties \cite{Lee.2016}. With respect to optoelectronic applications, semiconducting materials such as W- and Mo-based transition metal dichalcogenides (TMDCs) are especially appealing. These materials exhibit an indirect-to-direct band-gap transition upon reducing the thickness to a single layer (SL) \cite{Cheiwchanchamnangij.2012,Jin.2013,Cappelluti.2013,Zhang.2014,Yeh.2015}. Since the SL material has no inversion symmetry, the Kramers degeneracy is lifted which causes spin-dependent band splittings due to spin-orbit interaction. The spin texture with alternating spin orientations at the $\overline{\text{K}}$ and $\overline{\text{K}}$' high-symmetry points leads to coupled spin and valley physics and possible applications \cite{DiXiao.2012}. 

The valence bands of SL W- and Mo-based TMDCs have been studied in detail with photoemission techniques \cite{Klein.2001,Zhang.2014b,MaciejDendzik.2015,Miwa.2015,AntonijaGrubisicCabo.2015,DuyLe.2015,Zhang.2016,Bruix.2016,Ulstrup.2016,Ulstrup.2016,SungKwanMo.2016}. The detection of two different excitons A and B in optical experiments \cite{Mak.2010} is explained by the spin-dependent energy splitting of the valence band. So far, experimental information about the unoccupied conduction bands is limited to scanning tunneling spectroscopy \cite{Zhang.2015,Bruix.2016} and time-resolved photoemission data \cite{AntonijaGrubisicCabo.2015,Ulstrup.2016b,Ulstrup.2017}, yet without spin resolution. Since the conduction bands are also predicted to have a spin-dependent energy splitting \cite{Zhu.2011,Kosmider.2013,Liu.2013,Kormanyos.2015}, spin-allowed and spin-forbidden ("dark") transitions are expected. This leads to so-called dark excitons which possibly influence the efficiency of SL TMDC devices \cite{Ye.2015b,Echeverry.2016}. \

In this Letter, we use a combined angle-resolved photoelectron spectroscopy (ARPES) and angle-resolved  inverse-photoemission (IPE) setup, both with spin resolution, to investigate the spin texture of the highest valence bands and lowest conduction bands of SL WS$_2$ grown on Au(111). 
We compare our experimental results with band structure calculations for the isolated SL and for a SL on Au(111). 

	\begin{figure}[tbp!]
		\centering
		\includegraphics[width=0.90\columnwidth]{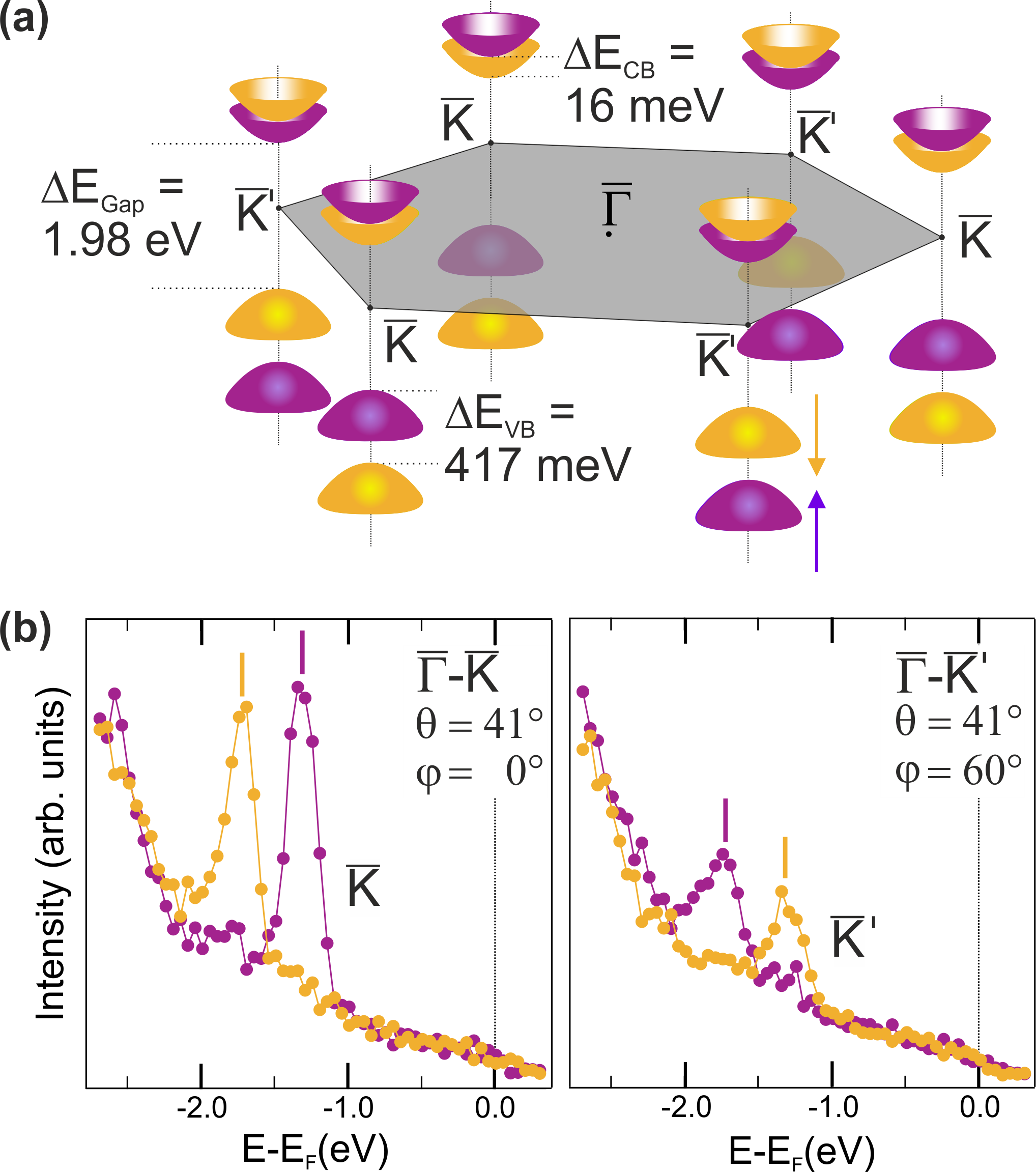}	
		\caption{(a) Sketch of the spin-dependent valley structure of the bands in single-layer WS$_2$ on Au(111) at the $\overline{\text{K}}$ and $\overline{\text{K}}$' high-symmetry points with a summary of the obtained results. (b) Spin-resolved ARPES spectra (out-of-plane spin sensitivity) of the uppermost valence bands at the $\overline{\text{K}}$ and $\overline{\text{K}}$' points. The spin splitting of the highest valence band $\Delta E_{\textnormal{VB}}$ is determined to $417\pm19$~meV.}
		\label{fig:1}
	\end{figure}

The sample used in the present work was single orientation, SL WS$_2$ on Au(111) with a coverage of about 45\%. For spin-resolved experiments, a single orientation of the SL WS$_2$ domains is essential because otherwise the measured spin polarization at the $\overline{\text{K}}$ and $\overline{\text{K}}$’ points would be reduced or even cancelled by mixed contributions from $\overline{\text{K}}$ and $\overline{\text{K}}$’. The SL WS$_2$ was grown at the SuperESCA beam line of the Elettra synchrotron-radiation facility in Trieste by evaporating tungsten at a partial pressure of H$_2$S onto the Au(111) substrate, the single orientation was verified by x-ray photoelectron diffraction (XPD) \cite{Bignardi.2018}. In Münster, the sample was annealed in ultrahigh vacuum to remove contaminants. The sample quality was checked with low-energy electron diffraction (LEED). The LEED pattern showed the expected moir\'{e} structure due to the lattice mismatch between WS$_2$ and Au as well as the same threefold rotational symmetry as measured just after sample preparation.

In our experimental approach, we are able to measure the energy dispersion and spin dependence of valence and conduction bands in the same chamber on the same sample \cite{Budke.2007}. ARPES measurements are performed with unpolarized light of a He discharge lamp (h$\nu=21.22$\,eV). The illuminated area on the sample is in the millimeter range. The photoemitted electrons are detected by a simulated 50 mm hemispherical analyzer (SHA 50 by FOCUS GmbH), that is mounted on a goniometer for angle-resolved measurements. The energy resolution was about 150 meV. The spin polarization of the emitted electrons is detected via spin-polarized low-energy electron diffraction (SPLEED) \cite{Feder.1981,Yu.2007}. The detector has a Sherman function of $S=0.24$ \cite{Budke.2007}. For spin-resolved IPE, we use a spin-polarized electron source, which provides an electron beam of 3 mm diameter with a spin polarization of $P=0.29$ \cite{Stolwijk.2014b} and a beam divergence of about $\pm2 ^\circ$ \cite{Zumbulte.2015}. For non-normal electron incidence on the sample, our setup is sensitive to the out-of-plane spin component \cite{Stolwijk.2014b}. Emitted photons of h$\nu=9.9$ eV are detected by a bandpass-type detector \cite{Budke.2007b,Thiede.2015,Thiede.2018}. The overall energy resolution of the IPE experiment is about 350 meV \cite{Budke.2007b}. The parallel component of the electron wave vector \textit{k$_{\parallel}$} is determined by the emission and incidence angles $\theta$ in ARPES and IPE, respectively. All spectra have been normalized to 100$\%$ Sherman function in ARPES and complete spin polarization of the incoming electrons in IPE \cite{Donath.1994}. During the measurements, the sample was at room temperature.

We have calculated the electronic structure of a freestanding WS$_2$ layer as well as a  SL WS$_2$ on a six-layer slab of Au(111) using density functional theory including spin-orbit coupling. The structural models and methods used for the calculations are described in detail in Ref. \cite{Dendzik.2017}.

\begin{figure}[tbp!]
		\centering
		\includegraphics[width=0.85\columnwidth]{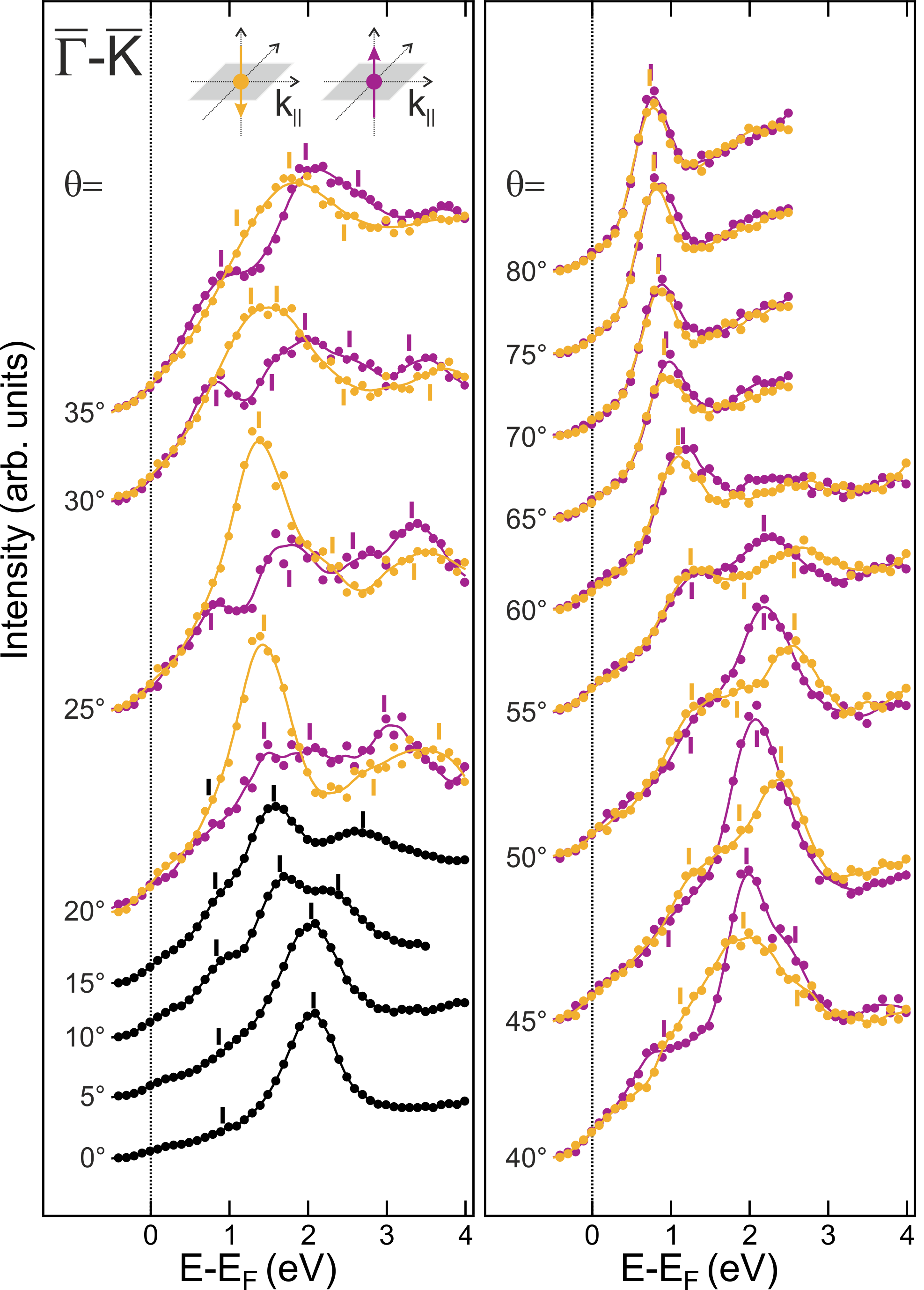}	
		\caption{Angle-resolved inverse photoemission spectra for single-layer WS$_2$ on Au(111) along $\overline{\Gamma}$-$\overline{\text{K}}$. For $\theta \leq 15^{\circ}$, spin-integrated data are shown as black dots.  For $\theta \geq 20^{\circ}$, spin-resolved data for out-of-plane spin sensitivity are presented as purple (yellow) dots for spin polarization parallel (antiparallel) to the surface normal. Vertical lines mark peak positions in the spectra. } 
		\label{fig:2}
	\end{figure}	
When studying the surface of bulk TMDC samples or SL films with multiple domain orientations, the experimental spin information is complicated by the signal of sublayers (spin-layer locking) \cite{Riley.2014,Razzoli.2017} or by mixed spin signals originating from $\overline{\text{K}}$ and $\overline{\text{K}}$' \cite{SungKwanMo.2016}, respectively. So far, a spin-resolved ARPES study on single-oriented SL TMDC is only reported for MoS$_2$ on Au(111), yielding out-of-plane spin-polarized valence bands with opposite sign at the $\overline{\text{K}}$ and $\overline{\text{K}}$' points \cite{Bana.2018}. An equivalent spin texture in the valence band is expected for WS$_2$ \cite{Zhu.2011,DiXiao.2012,JLi.2015,Tatsumi.2016}. Figure \ref{fig:1}(b) shows our spin-resolved ARPES measurements for SL WS$_2$/Au(111) at $\overline{\text{K}}$ and $\overline{\text{K}}$'. Purple and yellow dots denote data for electron spin polarization parallel or antiparallel to the surface normal, respectively. The valence band maximum (VBM) is found at \textit{E}$-E_{\textnormal{F}}=$ $-1.29\pm 0.02$~eV and $-1.71\pm 0.01$~eV for the two spin directions. Our results reveal a spin-dependent energy splitting $\Delta E_{\text{VB}}$ of $417\pm 19$~meV. This value is in good agreement with our calculation (431 meV) and other theoretical predictions \cite{Zhu.2011,Ramasubramaniam.2012,Kosmider.2013,Liu.2013,Ye.2014,Kormanyos.2015}. The same size of the splitting was obtained in spin-integrated measurements for SL WS$_2$ on different substrates \cite{MaciejDendzik.2015,Ulstrup.2016,Dendzik.2017} as well as on bulk samples \cite{Latzke.2015}, while other references report slightly higher values \cite{Yuan.2016,Forti.2017}.

Remarkably, the two oppositely spin-polarized valence band features (in the data for both $\overline{\text{K}}$ and $\overline{\text{K}}$') show almost 100\% spin polarization above background. Extrinsic spin-polarization effects caused by matrix-element effects based on orbital contributions as well as experimental parameters and geometry \cite{Donath.2018,Henk.2018} can be ruled out by the following experimental finding obtained with unpolarized light: in the same experimental geometry with only the sample rotated azimuthally by $\phi=60^{\circ}$, we obtain completely spin-polarized features at $\overline{\text{K}}$ and $\overline{\text{K}}$', yet with reversed sign. These observations are only possible if two conditions are met: (i) the SL film has one single orientation, thus confirming the XPD results \cite{Bignardi.2018}, and (ii) the bands at the Brillouin-zone boundary are intrinsically spin polarized.

To get information about the size of the energy gap and the spin dependence of the confining bands, IPE measurements of the conduction bands are necessary. Figure \ref{fig:2} presents IPE spectra for various angles of electron incidence $\theta$ along $\overline{\Gamma}$-$\overline{\text{K}}$. As mentioned before, in our setup, out-of-plane sensitivity is only available for $\theta \neq 0^{\circ}$ with increasing sensitivity for larger $\theta$. Therefore, spin-integrated data are shown as black dots for $\theta \leq 15^{\circ}$.  For $\theta \geq 20^{\circ}$, spin-resolved data for out-of-plane spin sensitivity are presented as purple (yellow) dots for spin polarization parallel (antiparallel) to the surface normal. Clear out-of-plane spin asymmetries in the conduction bands are detected. Estimated peak positions are marked by small vertical lines. The Fermi edge is visible in all ARPES (Fig. \ref{fig:1}) as well as IPE spectra (Fig. \ref{fig:2}). While WS$_2$ sustains its semiconducting properties when deposited on Au(111) \cite{Dendzik.2017}, the uncovered metallic Au(111) areas cause the Fermi level onset in the spectra.
 
\begin{figure}[tbp!]
		\centering
		\includegraphics[width=1\columnwidth]{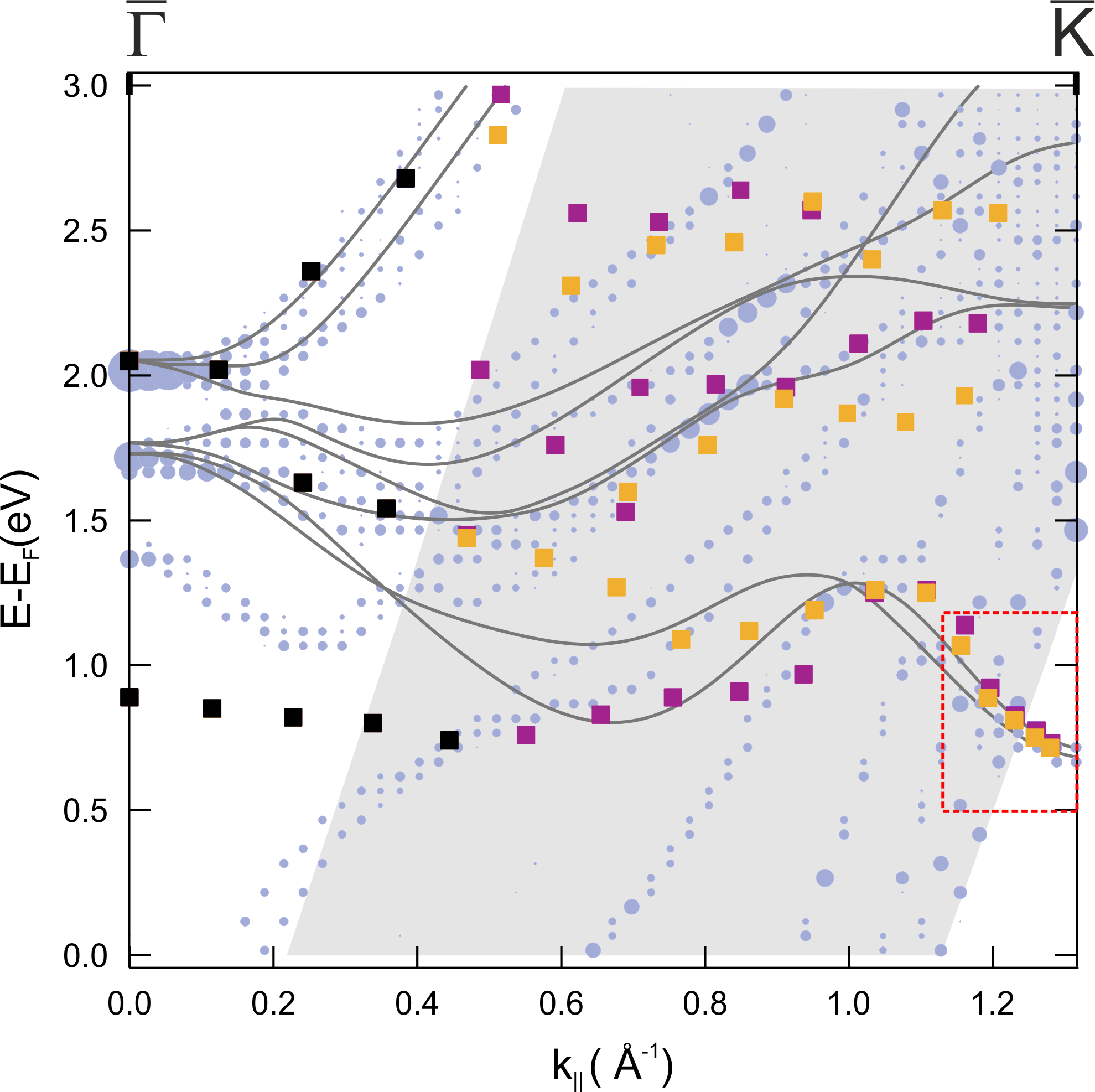}	
		\caption{E vs \textbf{k}$_{\parallel}$ band dispersions along $\overline{\Gamma}$-$\overline{\text{K}}$. Peak positions derived from the spectra in Fig. \ref{fig:2} are included as black, yellow, and purple squares for spin-integrated, out-of-plane spin-down and spin-up intensities, respectively. The grey-shaded area indicates bulk bands of Au, projected onto the (111) surface. Solid lines show results of a DFT calculation for a freestanding single layer of WS$_2$. Blue dots represent the band structure of WS$_2$ on Au(111) based on a DFT calculation. The region of the conduction bands near $\overline{\text{K}}$ is highlighted with a red box.} 
		\label{fig:3}
	\end{figure}

		Our experimental results for the conduction bands are summarized in an $E({\bf k}_\parallel)$ plot in Fig. \ref{fig:3}: Black, yellow, and purple squares denote peak positions in the spectra of Fig. \ref{fig:2} for spin-integrated, out-of-plane spin-down, and spin-up polarized data, respectively. The experimental data are presented along with calculations for (i) the projected bulk band structure of Au(111) [gray shaded area], (ii) a freestanding SL WS$_2$ [gray lines], and (iii) a SL WS$_2$ on top of a six layer slab Au(111) [blue dots]. The sizes of the blue dots are obtained from our supercell calculation and indicate the spectral weight of the effective band structure at each corresponding k point and energy interval resulting from the band unfolding method (for details, see \cite{Dendzik.2017}). The theoretical results have been rigidly shifted in energy to match the experimental results of the lowest conduction band at $\overline{\text{K}}$. Notably, with this calibration, the bands at $\overline \Gamma$ between 1.5 and 2.0 eV fit as well.
			
		Bands in regions where Au(111) has no states, e.g. close to $\overline \Gamma$ and below 1 eV close to $\overline{\text{K}}$, are expected to have almost pure WS$_2$ character. 
		Within the gray-shaded region Au bands exist, which might hybridize with WS$_2$ bands. The experimental energy dispersions follow predominantly the band dispersions of the freestanding layer with some deviations where bands of Au and WS$_2$ hybridize. The largest deviation between experiment and theory is observed for the lowest conduction band in the vicinity of $\overline \Gamma$, which appears in the experimental data with only low intensity. Remarkably, bands split off to lower energy around $\overline \Gamma$ are also theoretically expected for the adsorbed layer compared with the freestanding layer (see Fig. \ref{fig:3}), albeit not as much as experimentally observed.

An important question about SL TMDCs is the position of the conduction band minimum (CBM). Most studies indicate the CBM to be at the $\overline{\text{K}}$ point \cite{Klein.2001,Mak.2010,Splendiani.2010,Kumar.2012,Zhao.2013,Shi.2013,Roldan.2014,Sun.2016}. However, it is predicted that the energy at $\overline{\text{K}}$ is only few milli-electron-volts lower than at the so-called Q point about halfway between $\overline{\Gamma}$ and $\overline{\text{K}}$. There are even indications of the CBM being at the Q point \cite{Hsu.2017}. In our experiment, we find spectral features around Q, which are possibly influenced by Au states, with similar energies as the lowest spectral features at $\overline{\text{K}}$. Therefore, we cannot resolve whether the CBM position is at $\overline{\text{K}}$ or at Q. 

\begin{figure}[t!]
	\centering
	\includegraphics[width=1.0\columnwidth]{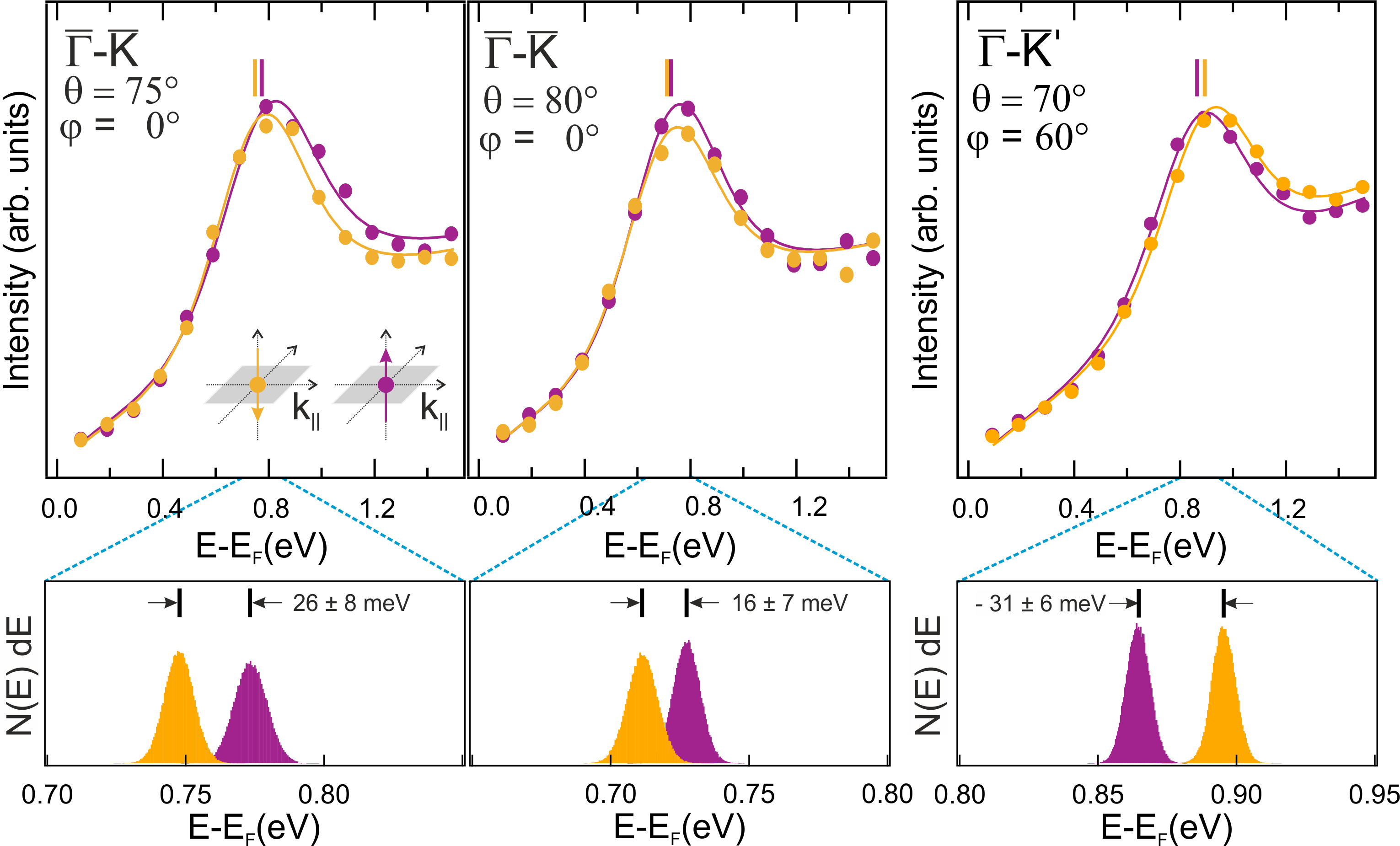}	
	\caption{Spin- and angle-resolved inverse photoemission spectra for out-of-plane spin sensitivity in the vicinity of the $\overline{\text{K}}$ and $\overline{\text{K}}$' points. Vertical lines mark the spin-dependent peak positions of the lowest conduction band. The lower panels show peak position distributions $N(E)dE$ of the conduction band emissions, obtained from least-squares fitting procedures (see text for details). Solid lines indicate the fit functions.}
	\label{fig:4}
\end{figure}

The key question with respect to the $\overline{\text{K}}$/$\overline{\text{K}}'$ valleys is the size of the energy gap and its spin structure. The lowest conduction band of WS$_2$ at $\overline{\text{K}}$ is found in a projected band gap of Au(111), and thus not influenced by Au states. Due to our photon energy of 9.9~eV, the accessible  ${\bf k}_\parallel$ range is limited. Nevertheless, our data for $\theta=80^\circ$ come very close to $\overline{\text{K}}$ at the given final-state energy (97$\%$ of $\overline{\Gamma}$-$\overline{\text{K}}$, see Fig. \ref{fig:3}). Figure \ref{fig:4} shows close-ups of spin-resolved IPE data for $\theta$ = $75^{\circ}$ and $\theta$ = $80^{\circ}$ along $\overline{\Gamma}$-$\overline{\text{K}}$ as well as data close to $\overline{\text{K}}$' ($\theta$ = $70^{\circ}$) to check the sign reversal of the spin signal.

The peak positions for spin-up and spin-down differ only slightly. Since the two partial spin spectra are measured separately, spin splittings can be resolved that are much smaller than the energy resolution or the intrinsic linewidth of the spectral features. In the case of completely spin-polarized states, the spin splitting can be determined quantitatively even in the case of energetically overlapping states. Otherwise, the obtained value is at least a lower limit. Based on our results for the valence bands, we reasonably assume that the conduction bands are completely spin polarized as well. 

We determined the peak positions of spin-up and spin-down spectra separately by a least-squares fitting procedure (see the Supplemntal Material for details. The fit function is composed of a Lorentzian function, a linear background, and a step function at the position of the Lorentzian function to simulate the 
steplike background increase due to secondary processes \cite{Dose.1980}. The result is then multiplied by the Fermi function and convoluted with a Gaussian-shaped apparatus function \cite{Budke.2007b,Stolwijk.2014b}. To quantify spin splittings and illustrate the statistical uncertainties, we used an approach reported earlier \cite{Passek.1992,Donath.1992}. For each spectrum, we generated a series of 100,000 pseudoexperimental spectra by varying each measured data point according to its statistical uncertainty and fitted the peak positions of the spectra. We obtained a peak-position distribution $N(E)dE$ for each measured spectrum and derived from these the spin splittings between respective partial spin spectra (examples are shown on an enlarged energy scale in the lower parts of \ref{fig:4}). For $\theta \geq 60^{\circ}$, all spectra along $\overline{\Gamma}$-$\overline{\text{K}}$ exhibit a spin splitting with the same sign (see Fig. \ref{fig:2}). An important test is the measurement on the sample rotated azimuthally by $60^{\circ}$: The data for $\theta = 70^{\circ}$ along $\overline{\Gamma}$-$\overline{\text{K}}$' also show a clear spin splitting, yet with reversed sign (see the right panel of Fig. \ref{fig:4})

The extracted spin splittings are a few tens of milli-electron-volts, decreasing to 31$\pm6$, 26$\pm8$ and 16$\pm7$ meV upon approaching the zone boundary for electron incidence angles of 70, 75, and 80$^\circ$, respectively. Our experimental value of $\Delta E_{\textnormal{CB}}=$ 16$\pm7$ meV is slightly lower than calculated conduction band splittings. We obtained 29 meV in good agreement with other calculations (26 to 32 meV \cite{Zhu.2011,Kosmider.2013,Liu.2013,Kormanyos.2015}).

The energy of the lowest conduction band in proximity to $\overline{\text{K}}$ is determined to $0.71\pm 0.03$~eV, which can be extrapolated to $0.69\pm 0.03$~eV at $\overline{\text{K}}$ by assuming a parabolic band behavior. Together with our result for the highest valence band, the size of the band gap amounts to $1.98\pm 0.04$~eV. Quasiparticle calculations predict the band gap of freestanding SL WS$_2$ in the range between 2.7 eV and 2.88 eV \cite{Ding.2011,Ramasubramaniam.2012,Liang.2013,Shi.2013,Kormanyos.2015}. Our determined band gap for WS$_2$/Au(111) is significantly lower due to the enhanced screening of the Au substrate, as reported also for MoS$_2$/Au(111) \cite{Bruix.2016}. Interestingly, a band gap of similar size (2.0 eV) was found for WS$_2$/Ag(111) by time-resolved ARPES \cite{Ulstrup.2017b}.

An essential piece of information is the spin sequence of the valence and conduction bands. Our data for WS$_2$ show that they are spin- split in the same way. In other words, the highest valence band is oppositely out-of-plane spin-polarized with respect to the lowest conduction band as sketched in Fig. \ref{fig:1} (a). Thus, the first spin-allowed (bright) transition is 16 meV higher in energy than the first dark transition.

Our experimental value for $\Delta E_{\textnormal{CB}}$ is important for theoretical studies, trying to determine the energy difference $\Delta E_{\text{Bright-Dark}}$ between dark and bright excitons \cite{Echeverry.2016}. Additionally, the electron-hole interaction within the exciton contributes to $\Delta E_{\text{Bright-Dark}}$. A few studies report on "brightening" the spin-forbidden dark excitons in WSe$_2$ \cite{Zhou.2017,Zhang.2017} and MoSe$_2$ \cite{JorgeQuereda.2018} by various methods. For SL WS$_2$, a splitting between dark and bright excitons $\Delta E_{\text{Bright-Dark}}= 47$ meV was reported from a photoluminescence experiment under the influence of an in-plane magnetic field \cite{MRMolas.2017}. While the lower energy of the dark exciton is consistent with our results, $\Delta E_{\text{Bright-Dark}}$ is much larger than $\Delta E_{\textnormal{CB}}$. For SL WS$_2$, the contribution of the electron-hole interaction is calculated to be in the order of 20 meV \cite{Echeverry.2016} partially explaining the difference between $\Delta E_{\textnormal{CB}}$ and $\Delta E_{\text{Bright-Dark}}$.
 
In conclusion, we studied the occupied and unoccupied electronic structure of SL WS$_2$/Au(111) experimentally by spin-resolved direct and inverse photoemission and theoretically by calculations for the freestanding SL and a SL adsorbed on Au(111). The total energy gap amounts to $1.98\pm 0.04$~eV, influenced by the screening of the Au substrate. Special attention was given to the spin structure of the VBM and the CBM at the $\overline{\text{K}}$/$\overline{\text{K}}$' valleys. Based on our results, we provide a schematic band structure at the $\overline{\text{K}}$ and $\overline{\text{K}}$' points, as it is sketched in Fig. \ref{fig:1}(a). The highest valence band is found to be spin-split by $417\pm19$~meV, the lowest conduction band by $16\pm7$~meV. The sequence of the spin-split bands is the same below and above the Fermi level; i.e., the highest valence band is oppositely out-of-plane spin-polarized with respect to the lowest conduction band. As a consequence, the lowest direct transition is spin-forbidden, i.e., optically dark. The first bright transition, involving the second conduction band, is 16 meV higher in energy than the band gap. Our calculations show that the bands at the $\overline{\text{K}}$/$\overline{\text{K}}$' valleys are almost unaffected by the Au substrate. Therefore, our results clarify important questions regarding band dispersion and spin structure for SL WS$_2$ with its promising valleytronic properties for future optoelectronic applications.

We thank Marco Bianchi, Jürgen Henk, Marcel Holtmann, Peter Krüger, and Paolo Lacovig for helpful support. This work was supported by the Danish Council for Independent Research, Natural Sciences under the Sapere Aude program (Grant No. DFF-4002-00029) and by VILLUM FONDEN via the Center of Excellence for Dirac Materials (Grant No. 11744).

\bibliography{Literature_180103}

  \end{document}